\documentclass[a4paper,11pt,twoside]{article}
\usepackage{amsmath,amsthm,amsfonts,amssymb,latexsym}
\usepackage{fancyhdr,graphicx}
\newtheorem{Theorem}{Theorem}

\newtheorem{Lemma}[Theorem]{Lemma}
\newtheorem{Remark}[Theorem]{Remark}

\newtheorem{Definition}[Theorem]{Definition}
\newtheorem{Notation}[Theorem]{Notation}

\begin{document}
\title{Some properties of the regular asynchronous systems}
\author{Serban E. Vlad}
\date{}
\maketitle

\fancypagestyle{plain} { \fancyhf{}
\fancyhead[LE]{Int. J. of Computers,
Communications and Control, Vol. 3 (2008)\\ Suppl. issue: Proceedings of ICCCC
2008,  pp. 11-16}
\fancyhead[RO]{Int. J. of Computers, Communications and
Control, Vol. 3 (2008)\\ Suppl. issue: Proceedings of ICCCC 2008,  pp. 11-16} }
\pagestyle{plain}

\renewcommand{\headrulewidth}{0.4pt}
\begin{abstract}
The asynchronous systems are the models of the asynchronous circuits from the digital electrical engineering. An asynchronous system $f$ is a
multi-valued function that assigns to each admissible input $u:\mathbf{R}%
\rightarrow \{0,1\}^{m}$ a set $f(u)$ of possible states $x\in f(u),x:%
\mathbf{R}\rightarrow \{0,1\}^{n}.$ A special case of asynchronous system
consists in the existence of a Boolean function $\Upsilon :\{0,1\}
^{n}\times \{0,1\}^{m}\rightarrow \{0,1\}^{n}$ such that $\forall
u,\forall x\in f(u),$ a certain equation involving $\Upsilon $ is fulfilled.
Then $\Upsilon $ is called the generator function of $f$ (Moisil used the
terminology of network function) and we say that $f$ is generated by $%
\Upsilon .$ The systems that have a generator function are called regular.

Our purpose is to continue the study of the generation of the asynchronous
systems that was started in \cite{bib5}, \cite{bib7}.
\end{abstract}

{\bf Keywords:} asynchronous system, regularity, generator function

\section{Preliminaries}

\begin{Notation}
Let be the arbitrary set $M$. The following notation will be useful: $%
P^{\ast }(M)=\{M^{\prime }|M^{\prime }\subset M,M^{\prime }\neq \emptyset
\}. $
\end{Notation}

\begin{Definition}
The set $\mathbf{B}=\{0,1\},$ endowed with the order $0\leq 1$ and with the
usual laws $\overline{\;\;},\cdot ,\cup ,\oplus $, is called the \textbf{%
binary Boole algebra}.
\end{Definition}

\begin{Definition}
The \textbf{%
initial value} $x(-\infty +0)\in \mathbf{B}^{n}$ of the function 
$x:\mathbf{R}\rightarrow \mathbf{B}^{n}$ is defined by $\exists
t^{\prime }\in \mathbf{R},\forall t<t^{\prime },x(t)=x(-\infty +0).$
\end{Definition}

\begin{Definition}
The \textbf{characteristic function} $\chi _{A}:\mathbf{R}\rightarrow 
\mathbf{B}$ of the set $A\subset \mathbf{R}$ is given by $\forall t\in 
\mathbf{R},\chi _{A}(t)=\left\{ 
\begin{array}{c}
1,t\in A \\ 
0,\;else%
\end{array}%
\right. .$
\end{Definition}

\begin{Notation}
We use the notation $Seq=\{(t_{k})|t_{k}\in \mathbf{R},k\in \mathbf{N},$ $%
t_{0}<...<t_{k}<...$ is unbounded from above$\}.$
\end{Notation}

\begin{Definition}
A function $x:\mathbf{R}\rightarrow \mathbf{B}^{n}$ is called $n-$\textbf{%
signal}, shortly \textbf{signal} if $\mu \in B^{n}$ and $(t_{k})\in Seq$
exist such that%
\begin{equation}
x(t)=\mu \cdot \chi _{(-\infty ,t_{0})}(t)\oplus x(t_{0})\cdot \chi
_{\lbrack t_{0},t_{1})}(t)\oplus ...\oplus x(t_{k})\cdot \chi _{\lbrack
t_{k},t_{k+1})}(t)\oplus ...  \label{pre1}
\end{equation}
The set of the $n-$signals is denoted by $S^{(n)}$.
\end{Definition}

\begin{Remark}
Let be $x:\mathbf{R}\rightarrow \mathbf{B}^{n},u:\mathbf{R}\rightarrow 
\mathbf{B}^{m}.$ Instead of $x\times u:\mathbf{R}\times \mathbf{R}%
\rightarrow \mathbf{B}^{n}\times \mathbf{B}^{m}$ we define the function $%
x\times u,$ many times denoted by $(x,u),$ as $x\times u:\mathbf{R}%
\rightarrow \mathbf{B}^{n}\times \mathbf{B}^{m}$ due to the existence of a
unique time variable $t\in \mathbf{R}$. Between the consequences derived from 
here we have the
identifications $S^{(n)}\times S^{(m)}=S^{(n+m)}$ and $P^{\ast
}(S^{(n)})\times P^{\ast }(S^{(m)})=P^{\ast }(S^{(n+m)}).$
\end{Remark}

\begin{Definition}
The \textbf{left limit} $x(t-0)$ of $x(t)$ from (\ref{pre1}) is the $\mathbf{%
R}\rightarrow \mathbf{B}^{n}$ function defined as $x(t-0)=\mu \cdot \chi
_{(-\infty ,t_{0}]}(t)\oplus x(t_{0})\cdot \chi _{(t_{0},t_{1}]}(t)\oplus
...\oplus x(t_{k})\cdot \chi _{(t_{k},t_{k+1}]}(t)\oplus ...$
\end{Definition}

\begin{Definition}
Let be $U\in P^{\ast }(S^{(m)}).$ A multi-valued function $f:U\rightarrow
P^{\ast }(S^{(n)})$ is called \textbf{asynchronous system}, shortly \textbf{%
system}. Any $u\in U$ is called (\textbf{admissible}) \textbf{input} and the
functions $x\in f(u)$ are called (\textbf{possible}) \textbf{states}.
\end{Definition}

\begin{Remark}
The asynchronous systems are the models of the asynchronous circuits. The
multi-valued character of the cause-effect association is due to the
statistical fluctuations in the fabrication process, the variations in the
ambiental temperature, the power supply etc. Sometimes the systems are given
by equations and/or inequalities.
\end{Remark}

\begin{Definition}
The \textbf{initial state function} of $f$ is by definition the function $%
i_{f}:U\rightarrow P^{\ast }(\mathbf{B}^{n}),\forall u\in
U,i_{f}(u)=\{x(-\infty +0)|x\in f(u)\}.$
\end{Definition}

\begin{Definition}
The function $\rho :\mathbf{R}\rightarrow \mathbf{B}^{n}$ is called \textbf{%
progressive} if $(t_{k})\in Seq$ exists such that $\rho (t)=\rho
(t_{0})\cdot \chi _{\{t_{0}\}}(t)\oplus ...\oplus \rho (t_{k})\cdot \chi
_{\{t_{k}\}}(t)\oplus ...$ and $\forall i\in \{1,...,n\},$ the set $\{k|k\in 
\mathbf{N},\rho _{i}(t_{k})=1\}$ is infinite. The set of the progressive
functions is denoted by $P_{n}.$
\end{Definition}

\begin{Notation}
Let be $\Upsilon :\mathbf{B}^{n}\times \mathbf{B}^{m}\rightarrow \mathbf{B}%
^{n},u\in S^{(m)},$ $\mu \in \mathbf{B}^{n}$ and $\rho \in P_{n}.$ The
solution of the equation%
\begin{equation}
\left\{ 
\begin{array}{c}
x(-\infty +0)=\mu \\ 
\forall i\in \{1,...,n\},x_{i}(t)=\left\{ 
\begin{array}{c}
\Upsilon _{i}(x(t-0),u(t-0)),if\;\rho _{i}(t)=1 \\ 
x_{i}(t-0),otherwise%
\end{array}%
\right.%
\end{array}%
\right.  \label{pre1_}
\end{equation}%
is denoted by $\Upsilon ^{-\rho }(t,\mu ,u).$
\end{Notation}

\begin{Definition}
The system $\Sigma _{\Upsilon }^{-}:S^{(m)}\rightarrow P^{\ast
}(S^{(n)}),\forall u\in S^{(m)},\Sigma _{\Upsilon }^{-}(u)=\{\Upsilon
^{-\rho }(t,\mu ,u)|\mu \in \mathbf{B}^{n},\rho \in P_{n}\}$ is called the 
\textbf{universal regular asynchronous system} that is generated by the
function $\Upsilon .$
\end{Definition}

\begin{Definition}
The system $f$ is called \textbf{regular} if $\Upsilon $ exists such that $%
\forall u\in U,f(u)\subset \Sigma _{\Upsilon }^{-}(u).$ If so, $\Upsilon $
is called the \textbf{generator function} of $f$ and we also say that $%
\Upsilon $ \textbf{generates} $f$.
\end{Definition}

\begin{Remark}
Equation (\ref{pre1_}) shows how the circuits compute asynchronously the
Boolean function $\Upsilon :$ the computation is made at the discrete time
instances $\{t_{k}|k\in \mathbf{N},\exists i\in \{1,...,n\},\rho
_{i}(t_{k})=1\}$ on these coordinates $\Upsilon _{i}$ for which $\rho
_{i}(t_{k})=1.$ The models of these circuits, the systems $f$ with the
generator function $\Upsilon ,$ have the remarkable property that a function 
$\pi _{f}:W_{f}\rightarrow P^{\ast }(P_{n})$ exists, $W_{f}=\{(x(-\infty
+0),u)|u\in U,x\in f(u)\}$ such that $\forall u\in U,f(u)=\{\Upsilon ^{-\rho
}(t,\mu ,u)|\mu \in i_{f}(u),\rho \in \pi _{f}(\mu ,u)\}.$ $\pi _{f}$ is
called the \textbf{computation function} of $f$. For $f$ regular, $\Upsilon $
and $\pi _{f}$ are not unique.
\end{Remark}

\section{Subsystems}

\begin{Definition}
The system $f$ is called a \textbf{subsystem} of $g:V\rightarrow P^{\ast
}(S^{(n)}),$ $V\in P^{\ast }(S^{(m)})$ and we write $f\subset g$, if $%
U\subset V$ and $\forall u\in U,f(u)\subset g(u).$
\end{Definition}

\begin{Remark}
We interpret $f\subset g$ in the following way: the systems $f$ and $g$
model the same circuit, but the model represented by $f$ is more precise
than the model represented by $g$.
\end{Remark}

\begin{Theorem}
The function $\Upsilon $ and the regular systems $f\subset \Sigma _{\Upsilon
}^{-},$ $g\subset \Sigma _{\Upsilon }^{-}$ are given. We denote by $%
i_{g}:V\rightarrow P^{\ast }(\mathbf{B}^{n})$ the initial state function and
by $\pi _{g}:W_{g}\rightarrow P^{\ast }(P_{n})$ the computation function of $%
g.$ The following statements are equivalent:

a) $f\subset g$

b) $U\subset V$ and $\forall u\in U,i_{f}(u)\subset i_{g}(u)$ and $\forall
u\in U,\forall \mu \in i_{f}(u),\forall \rho \in \pi _{f}(\mu ,u),\exists
\rho ^{\prime }\in \pi _{g}(\mu ,u),\Upsilon ^{-\rho }(t,\mu ,u)=\Upsilon
^{-\rho ^{\prime }}(t,\mu ,u).$
\end{Theorem}

\section{Dual systems}

\begin{Definition}
The \textbf{dual} function $\Upsilon ^{\ast }:\mathbf{B}^{n}\times \mathbf{B}%
^{m}\rightarrow \mathbf{B}^{n}$ of $\Upsilon $ is defined by $\forall (\mu
,\nu )\in \mathbf{B}^{n}\times \mathbf{B}^{m},\Upsilon ^{\ast }(\mu ,\nu )=%
\overline{\Upsilon (\overline{\mu },\overline{\nu })}.$ Here the bar $%
\overline{\mu }$ refers to the complement done coordinatewise.
\end{Definition}

\begin{Definition}
The \textbf{dual} of the system $f$ is by definition the system $f^{\ast
}:U^{\ast }\rightarrow P^{\ast }(S^{(n)}),$ where $U^{\ast }=\{\overline{u}%
|u\in U\}$ and $\forall u\in U^{\ast },f^{\ast }(u)=\{\overline{x}|x\in f(%
\overline{u})\}.$
\end{Definition}

\begin{Remark}
The system $f^{\ast }$ models the circuit modeled by $f$ with the AND gates
replaced by OR gates etc.
\end{Remark}

\begin{Notation}
We denote $i_{f^{\ast }}:U^{\ast }\rightarrow P^{\ast }(\mathbf{B}^{n})$, $%
\forall u\in U^{\ast },i_{f^{\ast }}(u)=\{\overline{\mu }|\mu \in i_{f}(%
\overline{u})\}.$
\end{Notation}

\begin{Notation}
We denote by $\pi _{f^{\ast }}:W_{f^{\ast }}\rightarrow P^{\ast }(P_{n})$
where $W_{f^{\ast }}=\{(\overline{x(-\infty +0)},u)|u\in U^{\ast },x\in f(%
\overline{u})\}$ the function $\forall (\mu ,u)\in W_{f^{\ast }},\pi
_{f^{\ast }}(\mu ,u)=\pi_{f}(\overline{\mu },\overline{u}).$
\end{Notation}

\begin{Theorem}
The dual system $f^{\ast }$ of $f\subset \Sigma _{\Upsilon }^{-}$ is
regular, $f^{\ast }\subset \Sigma _{\Upsilon ^{\ast }}^{-};$ its initial
state function is $i_{f^{\ast }}$ and its computation function is $\pi
_{f^{\ast }}.$
\end{Theorem}

\section{Cartesian product}

\begin{Definition}
The \textbf{Cartesian product} of the systems $f$ and $f^{\prime }:U^{\prime
}\rightarrow P^{\ast }(S^{(n^{\prime })}),U^{\prime }\in P^{\ast
}(S^{(m^{\prime })})$ is defined as $f\times f^{\prime }:U\times U^{\prime
}\rightarrow P^{\ast }(S^{(n+n^{\prime })}),$ $\forall (u,u^{\prime })\in
U\times U^{\prime },(f\times f^{\prime })(u,u^{\prime })=f(u)\times
f^{\prime }(u^{\prime }).$
\end{Definition}

\begin{Remark}
The Cartesian product $f\times f^{\prime }$ models two circuits that run
independently on each other.
\end{Remark}

\begin{Notation}
For $\Upsilon $ and $\Upsilon ^{\prime }:\mathbf{B}^{n^{\prime }}\times 
\mathbf{B}^{m^{\prime }}\rightarrow \mathbf{B}^{n^{\prime }}$, we denote by $%
\Upsilon \times \Upsilon ^{\prime }:\mathbf{B}^{n+n^{\prime }}\times \mathbf{%
B}^{m+m^{\prime }}\rightarrow \mathbf{B}^{n+n^{\prime }}$ the function $%
\forall ((\mu ,\mu ^{\prime }),(\nu ,\nu ^{\prime }))\in \mathbf{B}%
^{n+n^{\prime }}\times \mathbf{B}^{m+m^{\prime }},(\Upsilon \times \Upsilon
^{\prime })((\mu ,\mu ^{\prime }),(\nu ,\nu ^{\prime }))=(\Upsilon (\mu ,\nu
),\Upsilon ^{\prime }(\mu ^{\prime },\nu ^{\prime })).$ In this notation we
identify $(\mu ,\mu ^{\prime })\in \mathbf{B}^{n}\times \mathbf{B}%
^{n^{\prime }}$ with $(\mu _{1},...,\mu _{n},\mu _{1}^{\prime },...,\mu
_{n^{\prime }}^{\prime })\in \mathbf{B}^{n+n^{\prime }}$ etc.
\end{Notation}

\begin{Notation}
If $i_{f^{\prime }}:U^{\prime }\rightarrow P^{\ast }(\mathbf{B}^{n^{\prime
}})$ is the initial state function of $f^{\prime },$ we use the notation $%
i_{f\times f^{\prime }}:U\times U^{\prime }\rightarrow P^{\ast }(\mathbf{B}%
^{n+n^{\prime }}),\forall (u,u^{\prime })\in U\times U^{\prime },i_{f\times
f^{\prime }}(u,u^{\prime })=i_{f}(u)\times i_{f^{\prime }}(u^{\prime }).$
\end{Notation}

\begin{Notation}
The regular systems $f,$ $f^{\prime }$ are given, $f\subset \Sigma
_{\Upsilon },$ $f^{\prime }\subset \Sigma _{\Upsilon ^{\prime }}$ as well as
their computation functions: $\pi _{f}:W_{f}\rightarrow P^{\ast }(P_{n}),$ $%
\pi _{f^{\prime }}:W_{f^{\prime }}\rightarrow P^{\ast }(P_{n^{\prime }}).$
We denote by $\pi _{f\times f^{\prime }}:W_{f\times f^{\prime }}\rightarrow
P^{\ast }(P_{n+n^{\prime }})$ the function $W_{f\times f^{\prime
}}=\{((x(-\infty +0),x^{\prime }(-\infty +0)),(u,u^{\prime }))|(u,u^{\prime
})\in U\times U^{\prime },(x,x^{\prime })\in f(u)\times f^{\prime
}(u^{\prime })\},\forall ((\mu ,\mu ^{\prime }),(u,u^{\prime }))\in
W_{f\times f^{\prime }},$ $\pi _{f\times f^{\prime }}((\mu ,\mu ^{\prime
}),(u,u^{\prime }))=\pi _{f}(\mu ,u)\times \pi _{f^{\prime }}(\mu ^{\prime
},u^{\prime }).$
\end{Notation}

\begin{Remark}
The function $\pi _{f\times f^{\prime }}$ is correctly defined since $%
\forall \rho ,\forall \rho ^{\prime },\rho \in P_{n}$ and $\rho ^{\prime
}\in P_{n^{\prime }}\Longrightarrow (\rho ,\rho ^{\prime })\in
P_{n+n^{\prime }}.$
\end{Remark}

\begin{Theorem}
If $f\subset \Sigma _{\Upsilon }^{-},f^{\prime }\subset \Sigma _{\Upsilon
^{\prime }}^{-},$ then the system $f\times f^{\prime }$ is regular, $f\times
f^{\prime }\subset \Sigma _{\Upsilon \times \Upsilon ^{\prime }}^{-};$ its
initial state function is $i_{f\times f^{\prime }}$ and its computation
function is $\pi _{f\times f^{\prime }}.$
\end{Theorem}

\section{Parallel connection}

\begin{Definition}
The \textbf{parallel connection} of $f$ and $f_{1}^{\prime }:U_{1}^{\prime
}\rightarrow P^{\ast }(S^{(n^{\prime })}),$ $U_{1}^{\prime }\in P^{\ast
}(S^{(m)})$ is defined whenever $U\cap U_{1}^{\prime }\neq \emptyset $ by $%
f||f_{1}^{\prime }:U\cap U_{1}^{\prime }\rightarrow P^{\ast
}(S^{(n+n^{\prime })}),$ $\forall u\in U\cap U_{1}^{\prime
},(f||f_{1}^{\prime })(u)=f(u)\times f_{1}^{\prime }(u).$
\end{Definition}

\begin{Remark}
The parallel connection $f||f_{1}^{\prime }$ models two circuits that run
under the same input, independently on each other.
\end{Remark}

\begin{Notation}
Let be $\Upsilon $ and $\Upsilon _{1}^{\prime }:\mathbf{B}^{n^{\prime
}}\times \mathbf{B}^{m}\rightarrow \mathbf{B}^{n^{\prime }},$ for which we
denote by $\Upsilon ||\Upsilon _{1}^{\prime }:\mathbf{B}^{n+n^{\prime
}}\times \mathbf{B}^{m}\rightarrow \mathbf{B}^{n+n^{\prime }}$ the function $%
\forall ((\mu ,\mu ^{\prime }),\nu )\in \mathbf{B}^{n+n^{\prime }}\times 
\mathbf{B}^{m},(\Upsilon ||\Upsilon _{1}^{\prime })((\mu ,\mu ^{\prime
}),\nu )=(\Upsilon (\mu ,\nu ),\Upsilon _{1}^{\prime }(\mu ^{\prime },\nu
)). $
\end{Notation}

\begin{Notation}
Let $i_{f_{1}^{\prime }}:U_{1}^{\prime }\rightarrow P^{\ast }(\mathbf{B}%
^{n^{\prime }})$ be the initial state function of $f_{1}^{\prime }.$ If $%
U\cap U_{1}^{\prime }\neq \emptyset ,$ we use the notation $%
i_{f||f_{1}^{\prime }}:U\cap U_{1}^{\prime }\rightarrow P^{\ast }(\mathbf{B}%
^{n+n^{\prime }}),\forall u\in U\cap U_{1}^{\prime },i_{f||f_{1}^{\prime
}}(u)=i_{f}(u)\times i_{f_{1}^{\prime }}(u).$
\end{Notation}

\begin{Notation}
We suppose that the systems $f$, $f_{1}^{\prime }$ are regular i.e. $%
f\subset \Sigma _{\Upsilon }^{-},f_{1}^{\prime }\subset \Sigma _{\Upsilon
_{1}^{\prime }}^{-}$ and let $\pi _{f}:W_{f}\rightarrow P^{\ast }(P_{n}),\pi
_{f_{1}^{\prime }}:W_{f_{1}^{\prime }}\rightarrow P^{\ast }(P_{n^{\prime }})$
be their computation functions$.$ If $U\cap U_{1}^{\prime }\neq \emptyset ,$
then we use the notation $\pi _{f||f_{1}^{\prime }}:W_{f||f_{1}^{\prime
}}\rightarrow P^{\ast }(P_{n+n^{\prime }}),$ $W_{f||f_{1}^{\prime
}}=\{((x(-\infty +0),x^{\prime }(-\infty +0)),u)|u\in U\cap U_{1}^{\prime
},x\in f(u),x^{\prime }\in f_{1}^{\prime }(u)\},$ $\forall ((\mu ,\mu
^{\prime }),u)\in W_{f||f_{1}^{\prime }},\pi _{f||f_{1}^{\prime }}((\mu ,\mu
^{\prime }),u)=\pi _{f}(\mu ,u)\times \pi _{f_{1}^{\prime }}(\mu ^{\prime
},u).$
\end{Notation}

\begin{Theorem}
If $f\subset \Sigma _{\Upsilon }^{-},$ $f_{1}^{\prime }\subset \Sigma
_{\Upsilon _{1}^{\prime }}^{-}$ and $U\cap U_{1}^{\prime }\neq \emptyset ,$
then $f||f_{1}^{\prime }\subset \Sigma _{\Upsilon ||\Upsilon _{1}^{\prime
}}^{-};$ its initial state function is $i_{f||f_{1}^{\prime }}$ and its
computation function is $\pi _{f||f_{1}^{\prime }}.$
\end{Theorem}

\section{Serial connection}

\begin{Remark}
Let be the systems $f$ and $h:X\rightarrow P^{\ast }(S^{(p)}),X\in P^{\ast
}(S^{(n)}).$ When $\underset{u\in U}{\bigcup }f(u)\subset X,$ the serial
connection of $f$ and $h$ is defined by $h\circ f:U\rightarrow P^{\ast
}(S^{(p)}),\forall u\in U,(h\circ f)(u)=\underset{x\in f(u)}{\bigcup }h(x).$
If $f$ and $h$ are regular, this definition means that in the systems of
equations%
\begin{equation}
\left\{ 
\begin{array}{c}
x(-\infty +0)=\mu  \\ 
\forall i\in \{1,...,n\},x_{i}(t)=\left\{ 
\begin{array}{c}
\Upsilon _{i}(x(t-0),u(t-0)),if\;\rho _{i}(t)=1 \\ 
x_{i}(t-0),otherwise%
\end{array}%
\right. 
\end{array}%
\right. ,  \label{ser1}
\end{equation}%
\begin{equation}
\left\{ 
\begin{array}{c}
y(-\infty +0)=\lambda  \\ 
\forall j\in \{1,...,p\},y_{j}(t)=\left\{ 
\begin{array}{c}
\vartheta _{j}(y(t-0),x(t-0)),if\;\varpi _{j}(t)=1 \\ 
y_{j}(t-0),otherwise%
\end{array}%
\right. 
\end{array}%
\right.   \label{ser2}
\end{equation}%
where $u\in S^{(m)},x\in S^{(n)},y\in S^{(p)},\mu \in \mathbf{B}^{n},$ $%
\lambda \in \mathbf{B}^{p},$ $\rho \in P_{n},$ $\varpi \in P_{p},$ $\Upsilon
:\mathbf{B}^{n}\times \mathbf{B}^{m}\rightarrow \mathbf{B}^{n},$ $\vartheta :%
\mathbf{B}^{p}\times \mathbf{B}^{n}\rightarrow \mathbf{B}^{p}$ we eliminate $%
x$. Because this does not give any information of the regularity of $h\circ
f,$ we choose to work with a slightly different system from $h\circ f,$ for
which $x$ is not eliminated.
\end{Remark}

\begin{Notation}
If $f$ and $h$ fulfill $\underset{u\in U}{\bigcup }f(u)\subset X,$ then we
denote by $h\ast f:U\rightarrow P^{\ast }(S^{(n+p)})$ the system $\forall
u\in U,(h\ast f)(u)=\{(x,y)|x\in f(u),y\in h(x)\}.$
\end{Notation}

\begin{Notation}
The function $\vartheta \ast \Upsilon :\mathbf{B}^{n+p}\times \mathbf{B}%
^{m}\rightarrow \mathbf{B}^{n+p}$ is defined by $\forall ((\mu ,\lambda
),\nu )\in \mathbf{B}^{n+p}\times \mathbf{B}^{m},(\vartheta \ast \Upsilon
)((\mu ,\lambda ),\nu )=(\Upsilon (\mu ,\nu ),\vartheta (\lambda ,\Upsilon (\mu ,\nu ) )).$
\end{Notation}

\begin{Remark}
The point is that, instead of eliminating $x$ in (\ref{ser1}), (\ref{ser2})
as $h\circ f$ does, we can work with $h\ast f$ and with the equation%
\begin{equation*}
\left\{ 
\begin{array}{c}
z(-\infty +0)=(\mu ,\lambda ) \\ 
\forall k\in \{1,...,n+p\},z_{k}(t)=\left\{ 
\begin{array}{c}
(\vartheta \ast \Upsilon )_{k}(z(t-0),u(t-0)),if\;(\rho ,\varpi )_{k}(t)=1
\\ 
z_{k}(t-0),otherwise%
\end{array}%
\right. 
\end{array}%
\right. 
\end{equation*}%
where $z\in S^{(n+p)}.$
\end{Remark}

\begin{Notation}
For $i_{h}:X\rightarrow P^{\ast }(\mathbf{B}^{p})$ the initial state
function of $h,$ we denote by $i_{h\ast f}:U\rightarrow P^{\ast }(\mathbf{B}%
^{n+p})$ the function $\forall u\in U,i_{h\ast f}(u)=\{(\mu ,\lambda )|\mu
\in i_{f}(u),\lambda \in \underset{x\in f(u),x(-\infty +0)=\mu }{\bigcup }%
i_{h}(x)\}.$
\end{Notation}

\begin{Notation}
We suppose that $\pi _{h}:W_{h}\rightarrow P^{\ast }(P_{p})$ is the
computation function of $h$, $W_{h}=\{(y(-\infty +0),x)|x\in X,y\in h(x)\}.$
We denote by $\pi _{h\ast f}:W_{h\ast f}\rightarrow P^{\ast }(P_{n+p})$ the
function $W_{h\ast f}=\{((x(-\infty +0),y(-\infty +0)),u)|u\in U,x\in
f(u),y\in h(x)\},$ $\forall ((\mu ,\lambda ),u)\in W_{h\ast f},$ $\pi
_{h\ast f}((\mu ,\lambda ),u)=\{(\rho ,\varpi )|\rho \in \pi _{f}(\mu
,u),\varpi \in \underset{x\in f(u),x(-\infty +0)=\mu }{\bigcup }\pi
_{h}(\lambda ,x)\}.$
\end{Notation}

\begin{Theorem}
The systems $f$ and $h$ are given such that the inclusion $\underset{u\in U}{%
\bigcup }f(u)\subset X$ is true. If the regularity properties $f\subset
\Sigma _{\Upsilon }^{-},h\subset \Sigma _{\vartheta }^{-}$ hold, then $h\ast
f\subset \Sigma _{\vartheta \ast \Upsilon }^{-}$; the initial state function
of $h\ast f$ is $i_{h\ast f}$ and its computation function is $\pi _{h\ast
f}.$
\end{Theorem}

\section{Intersection}

\begin{Definition}
The \textbf{intersection} of $f:U\rightarrow P^{\ast }(S^{(n)})$ and $%
g:V\rightarrow P^{\ast }(S^{(n)}),U,V\in P^{\ast }(S^{(m)})$ is defined
whenever $\exists u\in U\cap V,f(u)\cap g(u)\neq \emptyset $ by $f\cap
g:W\rightarrow P^{\ast }(S^{(n)}),$ $W=\{u|u\in U\cap V,f(u)\cap g(u)\neq
\emptyset \},$ $\forall u\in W,(f\cap g)(u)=f(u)\cap g(u).$
\end{Definition}

\begin{Remark}
The intersection of two systems is a model that results by the simultaneous
validity of two compatible models.
\end{Remark}

\begin{Notation}
When $W\neq \emptyset ,$ we use the notation $i_{f\cap g}:W\rightarrow
P^{\ast }(\mathbf{B}^{n}),\forall u\in W,i_{f\cap g}(u)=i_{f}(u)\cap
i_{g}(u).$
\end{Notation}

\begin{Notation}
We consider the regular systems $f,g$ for which the generator function $%
\Upsilon :\mathbf{B}^{n}\times \mathbf{B}^{m}\rightarrow \mathbf{B}^{n}$ is
given such that $f\subset \Sigma _{\Upsilon }^{-},g\subset \Sigma _{\Upsilon
}^{-}.$ Their computation functions are $\pi _{f}:W_{f}\rightarrow P^{\ast
}(P_{n}),$ $\pi _{g}:W_{g}\rightarrow P^{\ast }(P_{n}).$ If the set $W$ is
non-empty, then we use the notation $\pi _{f\cap g}:W_{f\cap g}\rightarrow
P^{\ast }(P_{n})$ for the function that is defined by $W_{f\cap
g}=\{(x(-\infty +0),u)|u\in W,x\in f(u)\cap g(u)\},\forall (\mu ,u)\in
W_{f\cap g},$ $\pi _{f\cap g}(\mu ,u)=\{\rho |\rho \in \pi _{f}(\mu
,u),\exists \rho ^{\prime }\in \pi _{g}(\mu ,u),\Upsilon ^{-\rho }(t
,\mu ,u)=\Upsilon ^{-\rho ^{\prime }}(t ,\mu ,u)\}.$
\end{Notation}

\begin{Remark}
We remark the satisfaction of the following property of symmetry: $W_{f\cap
g}=W_{g\cap f}$ and $\forall (\mu ,u)\in W_{f\cap g},\forall \rho \in \pi
_{f\cap g}(\mu ,u),\exists \rho ^{\prime }\in \pi _{g\cap f}(\mu
,u),\Upsilon ^{-\rho }(t,\mu ,u)=\Upsilon ^{-\rho ^{\prime }}(t,\mu ,u)$ and 
$\forall \rho ^{\prime }\in \pi _{g\cap f}(\mu ,u),\exists \rho \in \pi
_{f\cap g}(\mu ,u),\Upsilon ^{-\rho ^{\prime }}(t,\mu ,u)=\Upsilon ^{-\rho
}(t,\mu ,u).$
\end{Remark}

\begin{Theorem}
If the regular systems $f\subset \Sigma _{\Upsilon }^{-},g\subset \Sigma
_{\Upsilon }^{-}$ fulfill $W\neq \emptyset ,$ then their intersection $f\cap
g:W\rightarrow P^{\ast }(S^{(n)})$ is regular $f\cap g\subset \Sigma
_{\Upsilon }^{-};$ its initial state function is $i_{f\cap g}$ and its
computation function is $\pi _{f\cap g}.$
\end{Theorem}

\section{Union}

\begin{Definition}
The \textbf{union} of $f,g$ is defined by $f\cup g:U\cup V\rightarrow
P^{\ast }(S^{(n)}),$ $\forall u\in U\cup V,(f\cup g)(u)=\left\{ 
\begin{array}{c}
f(u),u\in U\setminus V, \\ 
g(u),u\in V\setminus U, \\ 
f(u)\cup g(u),u\in U\cap V%
\end{array}%
\right. .$
\end{Definition}

\begin{Remark}
The union of the systems represents the validity of one of two models. This
is useful for example in testing, when $f$ is the model of the 'good'
circuit and $g$ is the model of the 'bad' circuit.
\end{Remark}

\begin{Notation}
We denote by $i_{f\cup g}:U\cup V\rightarrow P^{\ast }(\mathbf{B}^{n})$ the
function $\forall u\in U\cup V,i_{f\cup g}(u)=\left\{ 
\begin{array}{c}
i_{f}(u),u\in U\setminus V, \\ 
i_{g}(u),u\in V\setminus U, \\ 
i_{f}(u)\cup i_{g}(u),u\in U\cap V%
\end{array}%
\right. .$
\end{Notation}

\begin{Lemma}
The sets $W_{f}=\{(x(-\infty +0),u)|u\in U,x\in f(u)\},$ $W_{g}=\{(x(-\infty
+0),u)|u\in V,x\in g(u)\},$ $W_{f\cup g}=\{(x(-\infty +0),u)|u\in U\cup
V,x\in (f\cup g)(u)\}$ fulfill $W_{f\cup g}=W_{f}\cup W_{g}.$
\end{Lemma}

\begin{Notation}
Let be the regular systems $f,$ $g$ and the function $\Upsilon $ such that $%
f\subset \Sigma _{\Upsilon }^{-},g\subset \Sigma _{\Upsilon }^{-}$ are true$%
. $ The computation functions of $f,g$ are $\pi _{f},$ $\pi _{g}.$ We denote
by $\pi _{f\cup g}:W_{f\cup g}\rightarrow P^{\ast }(P_{n})$ the function $%
\forall (\mu ,u)\in W_{f\cup g},$ $\pi _{f\cup g}(\mu ,u)=\left\{ 
\begin{array}{c}
\pi _{f}(\mu ,u),(\mu ,u)\in W_{f}\setminus W_{g}, \\ 
\pi _{g}(\mu ,u),(\mu ,u)\in W_{g}\setminus W_{f}, \\ 
\pi _{f}(\mu ,u)\cup \pi _{g}(\mu ,u),(\mu ,u)\in W_{f}\cap W_{g}%
\end{array}%
\right. .$
\end{Notation}

\begin{Theorem}
If the systems $f,g$ are regular $f\subset \Sigma _{\Upsilon }^{-},g\subset
\Sigma _{\Upsilon }^{-},$ then the union $f\cup g:U\cup V\rightarrow P^{\ast
}(S^{(n)})$ is regular, $f\cup g\subset \Sigma _{\Upsilon }^{-};$ its
initial state function is $i_{f\cup g}$ and its computation function is $%
\pi _{f\cup g}.$
\end{Theorem}

%Signature
\begin{flushright}
Serban E. Vlad \\
Oradea City Hall \\
Computers Department \\
Piata Unirii, Nr. 1, 410100, Oradea, Romania \\
E-mail: serban\_e\_vlad@yahoo.com \\
Received: December 24, 2007 \\
\end{flushright}


\begin{thebibliography}{9}
\small

\bibitem{bib6}
S. E. Vlad, 
{\em Teoria sistemelor asincrone,}
Editura Pamantul Pitesti and WSEAS Press Athens, 2007.

\bibitem{bib5}
S. E. Vlad,
``Boolean dynamical systems'',
{\em the 15-th Conference on Applied and Industrial Mathematics CAIM 2007, Mioveni,
Romania, October 12-14}, 2007.

\bibitem{bib7}
S. E. Vlad,
``On the generation of the asynchronous systems'',
{\em the first National Conference of Applied and Fundamental
Mathematics, Iasi, Romania, November 9-10}, 2007.

\end{thebibliography}
\end{document}